\documentclass[aps,prl,twocolumn,superscriptaddress,showpacs]{revtex4}

\usepackage{graphicx}
\usepackage{epsfig}
\usepackage{natbib}
\begin{document}
\title{Shot noise spectrum of artificial Single Molecule Magnets: measuring spin-relaxation times via the Dicke effect.}
\author{L.D. Contreras-Pulido}
\affiliation{Departamento de Teor\'ia de la Materia Condensada,
Instituto de Ciencia de Materiales de Madrid, CSIC, Cantoblanco
28049, Madrid, Spain}

\author{R. Aguado}
\affiliation{Departamento de Teor\'ia de la Materia Condensada,
Instituto de Ciencia de Materiales de Madrid, CSIC, Cantoblanco
28049, Madrid, Spain}

\date{\today}

\begin{abstract}
We investigate theoretically shot noise in an artificial Single
Molecule Magnet based upon a CdTe quantum dot doped with a
\emph{single} S=5/2 Mn spin and gated in the hole sector. Mn-hole
exchange anisotropy is shown to lead to profound consequences on
noise, like super-Poissonian behaviour. We report on a novel effect, 
similar to the Dicke effect in Quantum Optics, that allows to
\emph{separately} extract the hole and Mn spin relaxation times
using frequency-resolved shot noise measurements.
We expect that our findings may have further relevance to
experiments in other $S>1/2$ systems including transport through
$Mn_{12}$ molecules and STM spectroscopy of magnetic atoms.
\end{abstract}

\maketitle
Single Molecule Magnets (SMMs) combine properties of a magnet with
those of a nanostructure. This makes SMMs a useful platform to merge
concepts and applications of spintronics and nanoelectronics
\cite{wernsdorfer08}. In a transistor
setup, SMMs behave as Quantum Dots (QDs). 
This allows detailed Coulomb Blockade (CB) level spectroscopy of
prototypical examples, like $Mn_{12}$\cite{heersche,jo} or
endofullerene $N@C_{60}$ \cite{grose}. The experimental extraction
of their intrinsic properties is, however, very challenging because
little is known about the precise role of various sample-preparation
steps like the attachment of leads. An alternative route is to study
artificial counterparts to a SMM, like a CdTe QD doped with a single
$\text{Mn}^{+2}$ ion with spin $S=5/2$ (Fig. 1). This proposal is
based on recent experiments where these QDs (without contacts) are
optically probed \cite{besombes1,besombes2,leger1,leger2,leger3}.
When doped with a single hole, the system behaves as a SMM with
magnetization steps and hysteresis \cite{joaquin-ramon1}.
Furthermore, its transport properties depend on the quantum state of
the Mn spin, giving rise to remarkable phenomena like hysteretic CB
\cite{joaquin-ramon2}.

While dc transport entails considerable information, a complete
understanding requires to go beyond and study shot noise
\cite{blanter}, which we address here. Our results show that
exchange anisotropy leads to super-Poissonionan shot noise due to
lack of relaxation of the Mn spin. Interestingly, a novel effect at
finite frequencies, similar to the Dicke effect in Quantum Optics
\cite{Dicke}, allows to \emph{separately} measure the hole and Mn
spin relaxation times (Fig. 4).

\paragraph{Model.---}The minimal model of an artificial SMM reads $H_{eff}^{QD}=\sum_{\sigma}\varepsilon_{h}d_{\sigma}^{\dagger}d_{\sigma}+H_{eff}^{exch}$.
The first term describes confined holes within the QD, the second
term describing the effective hole-Mn anisotropic exchange is
\begin{eqnarray}
H_{eff}^{exch}&=&[J_{||}\tau_h^zM^z+\frac{J_\perp}{2}(\tau_h^+M^-+\tau_h^-M^+)],
\label{Hdot}
\end{eqnarray}
with $J_{||}=J$ and $J_\perp=\alpha J$ ($\alpha\leq 1$). $M^{z,\pm}$
are the Mn spin operators and $\tau_h^{z,\pm}$ are Pauli matrices
operating in the hole lowest energy doublet
$|\Uparrow_h\rangle=a|3/2,+3/2\rangle+b|3/2,-1/2\rangle$ and
$|\Downarrow_h\rangle=a|3/2,-3/2\rangle+b|3/2,+1/2\rangle$.
$H_{eff}^{QD}$ is obtained from the full problem
$H=\sum_{n}\varepsilon_{n}d_{n}^{\dagger}d_{n}+J_h\vec{S}_h(\vec{r}_M)\vec{M}$
after projecting onto the $|\Uparrow_h\rangle$,
$|\Downarrow_h\rangle$ subspace \cite{joaquin-ramon2}.
$d_{n}^{\dagger}$ creates a confined hole in the spin-orbital
$\psi_{n}(\vec{r})$, described by a six band Konh-Luttinger
Hamiltonian, and $\vec{S}_h(\vec{r}_M)$ is the hole spin density
evaluated at the Mn location. $J=J_h|\psi(\vec{r}_M)|^2$, $J_h\simeq
60 eV \AA^3$ being the hole-Mn exchange coupling constant of CdTe.
Due to strong spin-orbit coupling, exchange is highly anisotropic
and spin flips are suppressed $\langle
\Downarrow_h|S_h^-|\Uparrow_h\rangle\approx 0$ unless there is some
heavy hole-light hole mixing due to asymmetry $b\neq 0$,
$\alpha=J_\perp/J\neq 0$ \cite{Korringa,alpha1}.
\begin{figure}[tb]
 \begin{center}
   \includegraphics[angle=0,width=.3\textwidth]{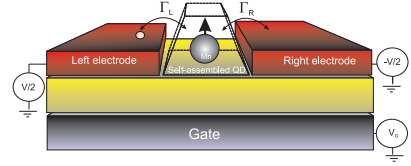}
\caption{(Color online) Schematics of the artificial SMM based on a
CdTe QD doped with a single $\text{Mn}^{+2}$ ion and coupled to
metallic reservoirs (tunnelling rates $\Gamma_L$ and $\Gamma_R$ and
chemical potentials $\mu_{L/R}=\pm V/2$). $V_G$ is the gate
voltage.}
    \label{diagram}
  \end{center}
\end{figure}
\begin{table}[t]\footnotesize
\begin{tabular}{|l|l|}
\hline
N=0 & \\
$|-5/2,0 \rangle$,$|+5/2,0 \rangle$  & $E=0$\\
$|-3/2,0 \rangle$,$|+3/2,0 \rangle$ & $E=0$\\
$|-1/2,0 \rangle$,$|+ 1/2,0 \rangle$ & $E=0$\\
\hline
N=1, antiferromagnetic (AF)& \\
$|-5/2,\Uparrow_h \rangle$,$|+5/2,\Downarrow_h \rangle$ & $E=\varepsilon_h-5/4 J$\\
$|-3/2,\Uparrow_h \rangle$,$|+3/2,\Downarrow_h \rangle$& $E=\varepsilon_h-3/4 J$\\
$|-1/2,\Uparrow_h \rangle$,$|+ 1/2,\Downarrow_h \rangle$ & $E=\varepsilon_h-1/4 J$\\
\hline
N=1, ferromagnetic (FM)& \\
$|-1/2,\Downarrow_h \rangle$,$|+ 1/2,\Uparrow_h \rangle$ & $E=\varepsilon_h+1/4 J$\\
$|-3/2,\Downarrow_h \rangle$,$|+3/2,\Uparrow_h \rangle$ & $E=\varepsilon_h+3/4 J$\\
$|-5/2,\Downarrow_h \rangle$,$|+5/2,\Uparrow_h \rangle$ & $E=\varepsilon_h+5/4 J$\\
\hline
\end{tabular}
\caption{States for the pure Ising case  (Eq. (\ref{Hdot}) with
$J_\perp=0$). N=0: the six $M_z$ projections of the Mn spin are
degenerate. N=1: the presence of a single hole breaks the
degeneracy. } \label{table1}
\end{table}

\paragraph{Quantum Master Equation (QME).---}The full Hamiltonian including
the coupling to reservoirs reads $H=H_{eff}^{QD}+\sum_{\alpha\in
{L,R}}H_{res}^\alpha+H_T^\alpha$. The Hamiltonian of each hole
reservoir is
$H_{res}^\alpha=\sum_{k_{\alpha},\sigma}\varepsilon_{k_{\alpha}\sigma}
c_{k_{\alpha}\sigma}^{\dagger}c_{k_{\alpha}\sigma}$ whereas
$H_T^{\alpha}=\sum_{k_{\alpha},\sigma}V_{k_{\alpha}}c_{k_{\alpha}\sigma}^{\dagger}d_{\sigma}+
H.c.$ describes tunneling. The QME for the reduced density matrix,
${\rho}(t)$, is obtained after applying a Born-Markov approximation
\cite{joaquin-ramon2} with respect to $H_T^{L/R}$,
$\dot{{\rho}}(t)={\mathcal L}{\rho}(t)$.
The dissipative dynamics is governed by a Liouvillean superoperator
${\mathcal L}$ which contains the forward/backwards ($\pm$)
transition rates $\Gamma^{\pm}_{N,N\mp
1}=\sum_{\alpha=L,R}\Gamma_\alpha f^{\pm}_{\alpha}(E_N-E_{N\mp
1})\sum_{\sigma}|\langle N|d^{\pm}_{\sigma}|N\mp 1\rangle|^2$
between eigenstates of $H_{eff}^{QD}$. The couplings $\Gamma_{L,R}$
are assumed to be constant and $f^+_\alpha$
($f^-_\alpha=1-f^+_\alpha$) is the Fermi function
\cite{joaquin-ramon2}. In the strong CB regime we only consider
states with N=0,1, holes (Table I).
The steady state ${\rho}^{stat}$, is obtained as
$\dot{{\rho}}(t)={\mathcal L}{\rho}^{stat}=0$, such that ${\mathcal
L}$ has a zero eigenvalue with right eigenvector
$|0\rangle\rangle\equiv\hat{\rho}^{stat}$.  The corresponding left
eigenvector is $|\tilde{0}\rangle\rangle\equiv\hat{1}$ such that
$\langle\langle
\tilde{0}|0\rangle\rangle=Tr[\hat{1}\hat{\rho}^{stat}]=1$
\cite{flindt0}, and averages
read
$\langle\hat{A}\rangle=Tr\{\hat{A}\rho^{stat}\}=\langle\langle\tilde{0}|\hat{A}|0\rangle\rangle$.

\paragraph{Shot Noise.---} The shot noise spectrum $S(\omega)=
2\int_{-\infty }^{\infty }d\tau e^{i\omega\tau}\langle\{\Delta
\hat{I}(t+\tau),\Delta \hat{I}(t)\}\rangle $ is defined as the
Fourier transform of the irreducible (cumulant) correlation function
$\langle\langle
\hat{I}(t+\tau)\hat{I}(t)\rangle\rangle\equiv\langle\Delta
\hat{I}(t+\tau)\Delta\hat{I}(t)\rangle$, where $\Delta
\hat{I}(t)=\hat{I}(t)-\langle I\rangle$ measures deviations away
from the steady state current $\langle I\rangle$. With the help of
the MacDonalds formula \cite{MacDonald}, $S(\omega)$ can be
rewritten as $S(\omega)=\omega\int_{0}^{\infty}dt\sin{(\omega
t)}\frac{d}{dt}\langle\langle n^2(t)\rangle\rangle$. $\langle\langle
n^2\rangle\rangle$ is the variance of the number $n$ of holes being
transferred in, say, the right lead in the time interval t. This
process is stochastic and is governed by the probability
distribution $P_n(t)$ which gives all moments (or cumulants) like,
in particular, $\langle\langle n^2\rangle\rangle$.
In this context, $P_n(t)$ is expressed in terms of the so-called
$n$-resolved density operator, ${\rho}^{(n)}(t)$, as
$P_n(t)=Tr_{sys}\{{\rho}^{n}(t)\}$ \cite{plenio98b}.
${\rho}^{(n)}(t)$ fulfills a generalized QME
$\dot{{\rho}}^{(n)} =({\mathcal L-\mathcal L^+_R-\mathcal
L^-_R}){\rho}^{(n)}+{\mathcal L^+_R}{\rho}^{(n-1)}+{\mathcal
L^-_R}{\rho}^{(n+1)}$.
The superoperator ${\mathcal L^{\pm}_R}$ contains forward/backwards
rates. Using the above method, the noise at the right barrier reads:
\begin{eqnarray}
 S_{RR}(\omega)=\langle\langle
\tilde{0}|{\mathcal J}|0\rangle\rangle-\langle\langle
\tilde{0}|{\mathcal I}R(\omega){\mathcal
I}|0\rangle\rangle-(\omega\rightarrow -\omega),
\label{srr}
\end{eqnarray}
${\mathcal I}\equiv{\mathcal L^+_R-\mathcal L^-_R}$ is the current
superoperator (such that $\langle I \rangle=\langle\langle
\tilde{0}|{\mathcal I}|0\rangle\rangle$) whereas ${\mathcal
J}\equiv{\mathcal L^+_R+ \mathcal L^-_R}$ is a superoperator
describing self-correlations at the barrier \cite{David}. The
pseudoinverse operator is defined as $R(\omega)\equiv
Q\frac{1}{i\omega-{\mathcal L}}Q$, where
$Q=1-|0\rangle\rangle\langle \tilde{0}|$ projects out the null space
defined by $|0\rangle\rangle\equiv\hat{\rho}^{stat}$ \cite{flindt0}.
Taking into account the left barrier $S_{LL}(\omega)$ and the noise
coming from charge accumulation \cite{aguado-brandes}, described by
the cross-correlations $S_{LR}(\omega),S_{RL}(\omega)$
\cite{lambert}, the total noise ($\Gamma_L=\Gamma_R$) reads
$S(\omega)=\frac{1}{4}\{S_{LL}(\omega)+S_{RR}(\omega)+S_{LR}(\omega)+S_{RL}(\omega)\}$.

\begin{figure*}
\begin{center}
\includegraphics[scale=0.5]{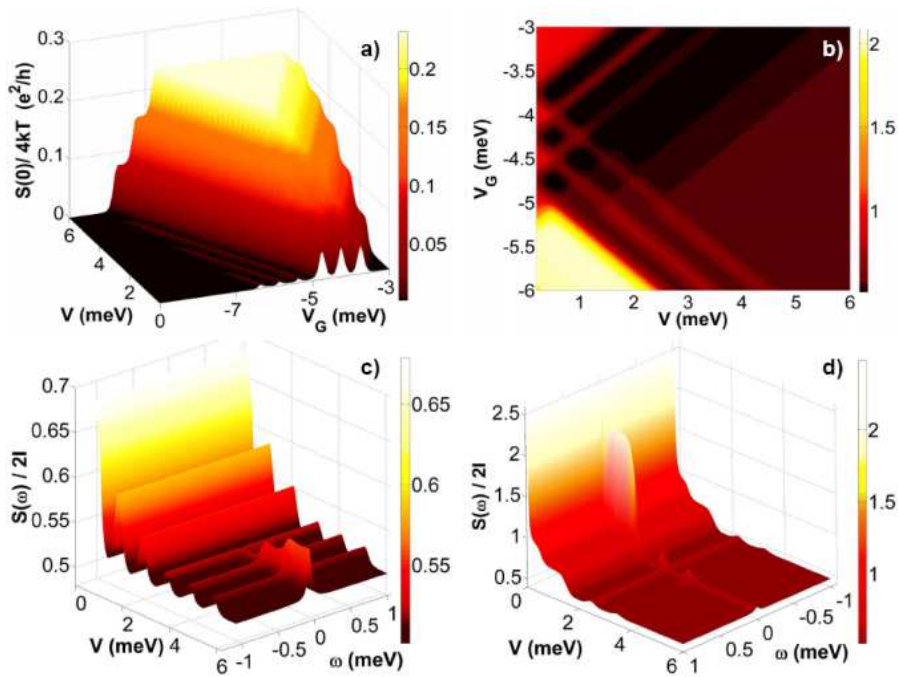}
\end{center}
\caption{(Color online) Noise of an artificial SMM for $\alpha=0$.
Top panels: Zero-frequency Shot noise (a) Fano factor (b). Bottom
panels: Finite frequency Fano factor, $S(\omega)/2\langle I\rangle$
for $V_G=-3.75$(c) and $V_G=-5.5$(d). Parameters: $\varepsilon_h=5$,
$J=1$, $\Gamma_L = \Gamma_R = 0.01$ and $T = 0.05$ (All energies in
meV)}
\end{figure*}

\paragraph{Results for $J_\perp=0$.---}The shot noise properties in the
$J_\perp=0$ case are investigated in Fig. 2. The top panels show
results for the $\omega=0$ shot noise (Fig. 2a) and Fano factor
$F=S(0)/2\langle I\rangle$ (Fig. 2b) as a function of $V$ and $V_G$.
At finite bias voltages, $S(0)$ presents steps which depend on
$V_G$. Between plateaus, $S(0)$ changes at values of $V$ where the
differential conductance $G_{dc}(V)=d\langle I\rangle/dV$ has maxima
(not shown) such that fluctuations are enhanced. In the limit
$V\rightarrow 0$, our calculation recovers the
fluctuation-dissipation theorem $S(0)=4kTG_{dc}(0)$, which relates
thermal fluctuations (Jonhson-Nyquist noise) with the linear
conductance. In this linear response regime, both shot noise and
linear conductance exhibit a three-peak structure in regions of gate
voltage corresponding to charge degeneracy between the $N=0$ and
$N=1$ sectors (from $V_G=-3$ to $V_G=-5$ in the figure). This is in
stark contrast with a normal QD which would exhibit instead one
single CB peak. Key for an understanding of this unusual CB is the
fact that $M_z$ does not relax during transport
\cite{joaquin-ramon2}: In the absence of holes, the spin of the Mn
ion is free and therefore all the six projections $M_z$ are
degenerate (Table I, top). Sweeping the gate voltage towards the
charge degeneracy region, the only allowed transitions are those
which conserve $M_z$. This results in three possible charge
degeneracy points corresponding to the transition
$N=0\Leftrightarrow N=1$ at different values of $M_z$. For example,
the first CB peak ($V_G=-3.75$) corresponds to charge degeneracy
between $|-5/2,0 \rangle$ and $|-5/2,\Uparrow_h \rangle$ (or $|5/2,0
\rangle$ and $|5/2,\Downarrow_h \rangle$). The charge degeneracy
condition for states with either $M_z=\pm 3/2$ or $M_z=\pm 1/2$
occurs at higher $|V_G|$ (Table I, middle) and, therefore CB is
spin-dependent.
By further decreasing $V_G$, we obtain a very small,
but finite, three-peak structure which can be attributed to thermal
excitations to excited states in the ferromagnetic (FM) sector
(Table I, bottom). The above physical picture leads to strong
spin-dependent deviations from Poissonian noise (Fano Factor, Fig.
2b). For $|V_G|\leq5$, the noise remains sub-Poissonian due to
spin-dependent CB. For $|V_G|>5$,
\begin{figure}[tb]
 \begin{center}
    \includegraphics[angle=0, width=0.45\textwidth]{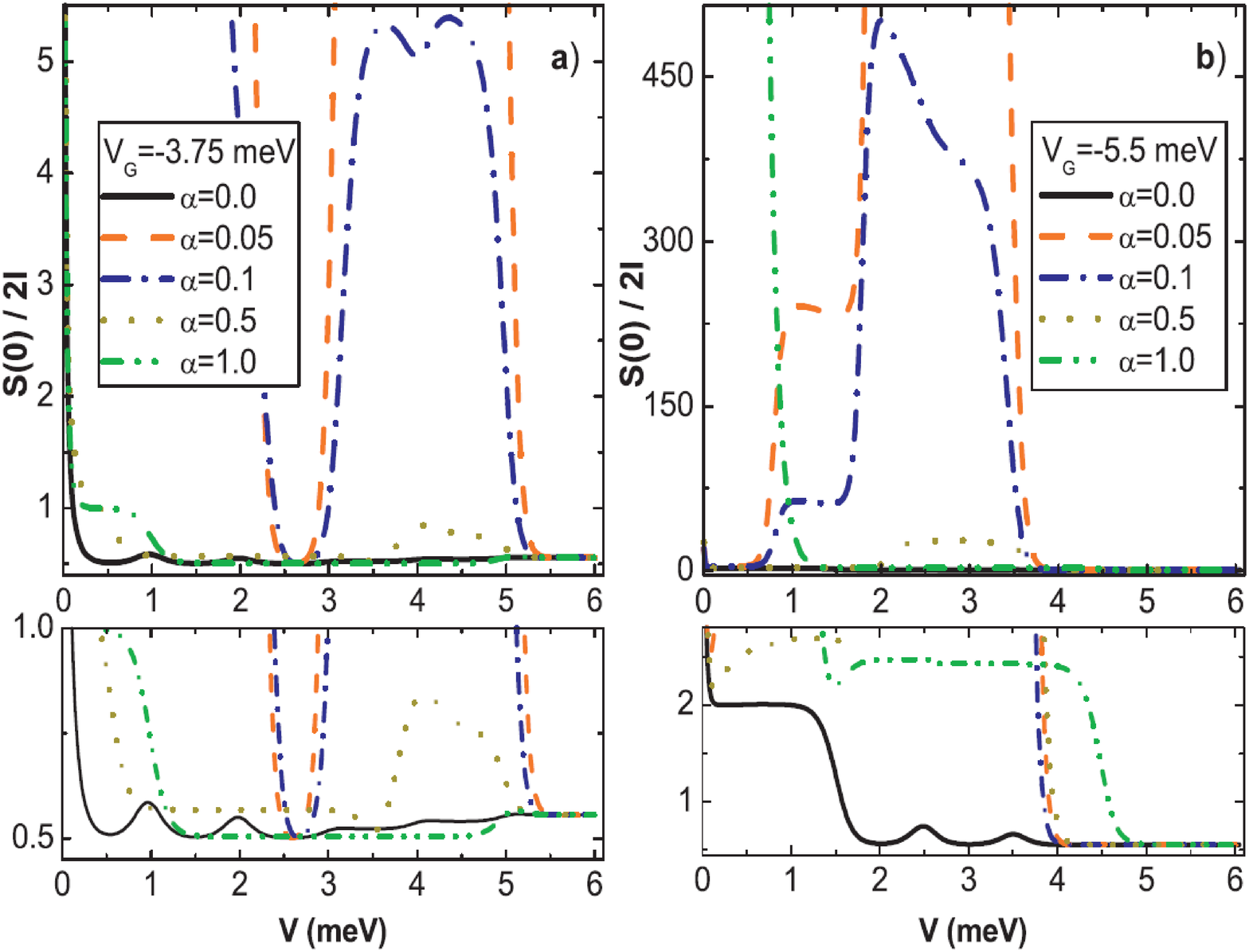}
    \caption{(Color online) Zero-frequency Fano factors for $\alpha\neq
    0$. a) $V_G=-3.75$, b) $V_G=-5.5$ (the rest of parameters are the same as in Fig. 2). Bottom panels: blow up of low Fano factor regions of the corresponding top panels.
    }
    \label{noise_curr}
  \end{center}
\end{figure}
the competing dynamics between slow (filled states in the AF sector)
and fast channels (FM sector) results in bunching and the noise
becomes super-Poissonian \cite{belzig1}. As $V$ increases, FM
excited states enter within the bias window and then the noise
becomes sub-Poissonian. In this case, bunching of holes at low
voltages is replaced by direct hole spin-flips due to tunneling
to/from the QD.
The results for finite $\omega$ are shown in the bottom panels of
Fig. 2. At $V_G=-3.75$ (Fig. 2c), shot noise is
frequency-independent up to $V\approx 3$. In this voltage region,
only states corresponding to the AF sector come into play.
Interestingly, at $V\geq 3$ a resonance around $\omega =0$ develops.
The appearance of this resonance can be understood as an enhancement
of fluctuations due to tunneling through both AF and FM channels
which are now within the bias window. For example, starting from the
state $|-1/2,\Uparrow_h \rangle$, one hole can tunnel out of the QD
leaving the Mn spin in the state $M_z=-1/2$. A second hole can now
tunnel onto the state $|-1/2,\Downarrow_h \rangle$, which is
energetically available, resulting in an effective spin relaxation
for the hole ($M_z$ does not change). This mechanism gives rise to a
resonance in $S(\omega)$ whose width \emph{is given by the inverse
hole spin-flip time $1/T_1^h\approx\Gamma$} (other intrinsic hole
spin-flip mechanisms, not included here, would also contribute to
this width). Further reduction of $V_G$ allows both kind of channels
to participate in transport even at low voltages and the resonance
is present for all $V$ (Fig. 2d), in good agreement with our
previous interpretation.

\paragraph{Role of spin-flip terms, $J_\perp\neq 0$.---} If valence band mixing is
nonzero, $|\pm 3/2\rangle$ heavy holes couple with $|\mp 1/2\rangle$
light holes ($J_{\perp}\neq 0$ in Eq. (\ref{Hdot}) and $b\neq 0$ in
the expressions for $|\Uparrow_h\rangle$ and
$|\Downarrow_h\rangle$). This mixing allows simultaneous spin flips
between the hole and the Mn spins \cite{leger2,joaquin} which
results in split states which are bonding and antibonding
combinations of $|M_z=-1/2, \Uparrow_h \rangle$ and $|M_z=+
1/2,\Downarrow_h\rangle$. This splitting can be extracted directly
from the $d\langle I\rangle/dV$ curves (not shown here). Fig. 3
illustrates the dramatic changes that mixing induces on noise. As
$\alpha=J_{\perp}/J$ increases, the Fano factor can reach values
$F>>1$ in voltage regions (both gate and bias) where the noise is
sub-Poissonian for $\alpha=0$ (Fig. 3a). Interestingly, the largest
Fano factors are obtained for small values of $\alpha$:
Most of the time, transport events conserve $M_z$. However,
spin-flips mediated by hole transport are small but nonzero. As a
result, periods of small current, followed by larger currents, are
possible leading to huge Fano factors. We emphasize that
super-Poissonian behavior is related here with Mn spin-flips. This
is in contrast with the $\alpha=0$ case where only hole spins can
flip. Although we do not seek here a detailed explanation of the
various $F>>1$ regions for different values of $V_G$ and $V$, we
mention in passing that each particular feature in Figs. 3 can be
associated with different transitions between $\alpha\neq 0$ states.
\paragraph{$S(\omega)$ and the Dicke effect.---}After many transport cycles occur, the Mn spin
relaxes completely in some typical time scale $T_1^M>>T_1^h$:
Starting from an empty QD with the Mn in the spin state $M_z$, a
hole tunnels and exchanges one unit of spin with the Mn. After some
time $t\approx 1/\Gamma$, the hole tunnels out of the QD such that
the Mn spin is now $M_z\pm$1. Note that it takes many tunneling
events to completely relax the Mn spin, such that $T_1^M>>T_1^h$.
This separation of time scales is reflected in $S(\omega)$, which,
remarkably, consists of a narrow peak on top of broad resonance
(Fig. 4a). The explanation of this feature is a process analogous to
the Dicke effect in Quantum Optics \cite{Dicke,chudnoski}: the
coupling of both relaxation channels (fast spin hole relaxation and
slow Mn spin relaxation) leads to a splitting into two combined
decay channels for the whole system. The width of the superradiant
channel (broad resonance) allows to extract the hole spin relaxation
time as $1/T_1^h$. More importantly, the subradiant channel (narrow
resonance) can be used to extract the intrinsic Mn spin-flip time as
$1/T_1^M$ \emph{in an independent manner}. When $\alpha=0$, only
hole spin relaxation is possible (Fig. 4a, solid line). As $\alpha$
changes, a extremely narrow resonance develops on top of the broad
one. Furthermore, the difference in magnitude between the noise
values at $\omega=0$ (see previous section) and
$1/T_1^h>\omega>1/T_1^M$ may facilitate the experimental
verification of this effect.
The broad resonances for $\alpha=0$ and $\alpha\neq 0$ are very
similar, as one expects from our interpretation in terms of hole
relaxation. When one of the spin relaxation channels is not
available, the Dicke effect should disappear. We have made
systematic checks, as a function of $V$, $V_G$ and $\alpha$, which
corroborate this and demonstrate the robustness of the effect. Fig.
2c, gives a clear demonstration of the absence of Dicke effect when
$\alpha=0$. The opposite case, namely when hole spin relaxation
starts to be inefficient, should result in the disappearance of the
broad feature in the noise (Fig.4b).
\begin{figure}[tb]
 \begin{center}
    \includegraphics[angle=0, width=0.29\textwidth]{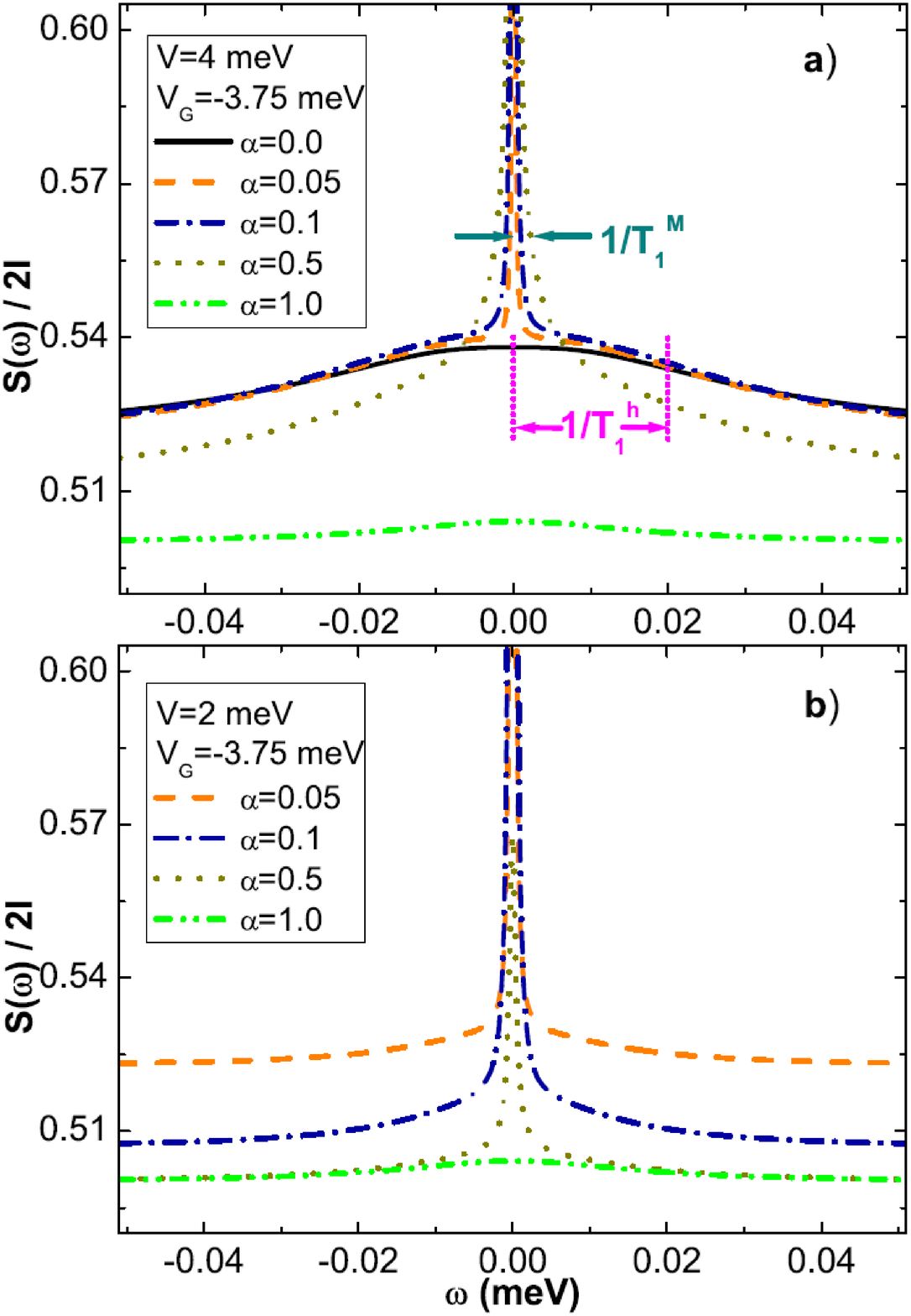}
    \caption{(Color online) a) Dicke effect in the shot noise spectrum $S(\omega)$ of an artificial
    SMM. The width of the superradiant channel (broad
    resonance) allows to extract the spin hole relaxation time as $1/T_1^h$
    whereas the subradiant channel (narrow resonance) gives the Mn spin relaxation time as
    $1/T_1^M$. b) At lower voltages, hole spin relaxation mediated
    by charge fluctuations is inefficient resulting in a single
    narrow peak due to Mn spin relaxation. For clarity, we do not show the
largest values at $\omega=0$, see main text.
    }
    \label{noise_curr}
  \end{center}
\end{figure}

\paragraph{Concluding remarks and experimental consequences.---} In
summary, our calculations show that Mn-hole exchange anisotropy has
profound consequences on shot noise, like super-Poissonionan
behavior due to incomplete relaxation of the Mn spin. The main
result of this Letter is a novel Dicke effect in the shot noise
spectrum that can be used to \emph{separately} measure the spin
relaxation times of, both, holes and Mn. Finally, most of the
physics captured by our model is inherent to large spin $S>1/2$
systems with anisotropy. We therefore expect that shot noise
measurements in other SMMs, like in the experiments of Refs.
[\onlinecite{heersche,jo,grose}], may reveal the effects described
here. STM spectroscopy of magnetic atoms such as $S=5/2$ Mn on
$Cu_2N$ \cite{Hirjibehedin} is one further experimental example
where our findings may be relevant.

We thank Joaqu\'{\i}n Fern\'andez-Rossier for his input on the
Mn-doped QD model and many useful discussions. Research supported by
MEC-Spain (Grant No. MAT2006-03741), CSIC and CAM (Grant No.
CCG08-CSIC/MAT- 3775). D. C. acknowledges financial support from
Mexican CONACyT ("Consolidaci\'on de Grupos de Investigaci\'on"
Postdoctoral Scholarship).


\begin{thebibliography}{0}
\bibitem{Korringa}
Strain-induced anisotropy terms of the form $D(M^z)^2$ are very
small as compared to Mn-hole exhange. This anisotropy is, however,
important when the Korringa mechanism contributes to Mn spin
relaxation in the absence of carriers. See Ref.
[\onlinecite{leger3}].
\end{thebibliography}
\end{document}